\documentclass{article}
\usepackage[numbers]{natbib}
\usepackage{graphicx}
\usepackage[colorlinks=true,linkcolor=blue,urlcolor=blue,citecolor=blue]{hyperref}

\begin{document}


\title{Digging the topology of rock art in Northwestern Patagonia}

\author{
{\sc Fernando E. Vargas, Jos\'e L. Lanata},\\[2pt]
Instituto de Investigaciones en Diversidad Cultural \\y Procesos de Cambio (IIDYPCA, CONICET, UNRN)\\
Mitre 630, San Carlos de Bariloche, R\'{\i}o Negro, Argentina
\\[6pt]
{\sc Guillermo Abramson, Marcelo N. Kuperman,}\\[2pt]
Centro At\'omico Bariloche and Instituto Balseiro\\
Comisi\'on Nacional de Energ\'{\i}a At\'omica (CNEA)\\
Consejo Nacional de Investigaciones Cient\'{\i}ficas y T\'ecnicas (CONICET)\\
Av. E. Bustillo 9500, R8402AGP San Carlos de Bariloche, \\R\'{\i}o Negro, Argentina
\\[6pt]
{\sc and}\\[6pt]
{\sc Danae Fiore}\\[2pt]
Asociaci\'on de Investigaciones Antropol\'ogicas (AIA, CONICET, UBA)\\
B. Mitre 1131 7G, 1036 Buenos Aires, Argentina.
}

\maketitle

\begin{abstract}
{We present a study on the rock art of northern 
Patagonia based on  network analysis and communities detection. We unveil a significant 
aggregation of archaeological sites, linked by common rock art motifs that 
turn out to be consistent with their geographical distribution and 
archaeological background of hunter-gatherer stages of regional peopling and land use. 
This exploratory study will allow us to approach more accurately some 
social strategies of visual communication entailed by 
rock art motif distribution, in space and time. }
{}
{Published in Journal of Complex Networks, cmz033 (2019) \url{https://doi.org/10.1093/comnet/cnz033}.}
\end{abstract}

\section{Introduction}

The use of analytical methods derived from network science has undeniably hatched over the last 
de\-ca\-de, bringing together several disciplines as mathematics, physics, computer science, biology 
and social sciences. Interdisciplinary approaches to the study of past and present societies, from the point of view of complex systems, have started to dominate the tendency towards the creation of knowledge in anthropology \cite{hamilton2007,apicella2012,migliano2017} and archaeology \cite{kohler2012,knappett2013,brughmans2016,mills17,gjesfjeld2015}. Within this context, the application of analytical 
methods derived from graph theory and from social network analysis to the study of archaeological problems has been increasing at a fast step. More examples of this interdisciplinary interaction can be found listed in \cite{collar15} and \cite{brughmans2013}, where the 
authors trace a thorough description of the state of the art of social network analysis and show its evolution within the last 50 years. 

As stated in the review by Mills \cite{mills17}, the use of networks in archaeology is especially relevant for studies that look for relational patterns between the 
units or entities that compose the system under analysis. Their 
usefulness resides in the possibility of abstracting the most relevant features of 
a case study and represent them and their connections as a network 
\cite{collar15,caridi16,radivojevic18}. In this way, the information can be analyzed 
globally and locally in a systematic manner by revealing the topological properties of the 
underlying network and the emerging patterns encoding valuable pieces of information. In fact, the 
structure associated to any cultural process is  organized into networks of nodes and relational 
links that connect them. Once the abstraction exercise has been done, different networks can be constructed from the same 
archaeological data to explore  different aspects of the same phenomena. One of the most relevant contributions of the application of complex networks in archaeology is its 
capacity to study large sets of archaeological data in different spatial scales and archaeological processes through time. 


Within these premises, formal networks have been applied in different theoretical frames, to different archaeological contexts and also to a wide range of archaeological evidence. Such is the case of a recent work focusing on a site scale \cite{hodder2016} that analyzes, from an entanglement perspective, how different materials are related in the clay use of \c{C}atalh\"{o}y\"{u}k (Anatolia). In a similar line, two very recent works \cite{mazzucato2019,Ladefoged19} applied social networks and community detection to analyze the formation of groups and social integration. In the former, the authors also analyze the \c{C}atalh\"{o}y\"{u}k site, and study a vast set of archaeological data to find the dynamics of interconnectivity and cooperation among different settlements. In the latter, obsidian artifacts of northern New Zealand, to trace social affiliation of M\={a}ori people during a time of social changes.

Another significant approach is the one oriented to solve problems related to procurement, distribution and use of raw materials, like obsidian and 
pottery compositions \cite{gjesfjeld2015,golitko2015,gjesfjeld2013}. From a macroregional approach, a wide variety of works can be found in 
Knappett's compilation \cite{knappett2013}. Currently, a prominent work in this line of macroregional approaches is that of Radivojevic and Grujic 
\cite{radivojevic18} who, by means of community detection applied to a chemical data of copper-based objects, identify a significant community network 
of metal production and exchange during 3000 years in East-Central Europe.

Also, within this increasing context of network analysis in archaeology, there are studies that discuss the methodological and even the epistemological challenges of the application of networks in the archaeological field 
\cite{gjesfjeld2015,collar15,peeples2013}.

The application of formal network analysis in rock art studies do not have a long tradition and, in the literature, we found only a few cases 
\cite{caridi16,alexander2008,riris2019}.
The pioneering work of Alexander \cite{alexander2008} applied for the first time analytical methods to the study of a single panel of the rock art 
site known as the Bedolina Map, in Valcamonica (Italy). His goal was to shed light on the structural composition of the complex motif. Another 
relevant work is that of Caridi and Scheinsohn \cite{caridi16}, who built a mutual information network from a set of data of western Patagonian rock 
art, to the south of our study region. Their goal was to model possible paths of cultural and information transmission by means of rock art motifs, by 
identifying clusters of associated sites. Their results show five clusters with a strong overlapping, interpreted as different moments of cultural 
transmission paths. Finally, a recent paper by Riris and Oliver \cite{riris2019} applies network analysis to a data set of rock art in the Orinoco 
basin (Venezuela), to analyze the possible links or discontinuities within a set of sites in terms of stylistic attributes. 

\begin{figure}[t]
\centering
\includegraphics[width=0.8\columnwidth]{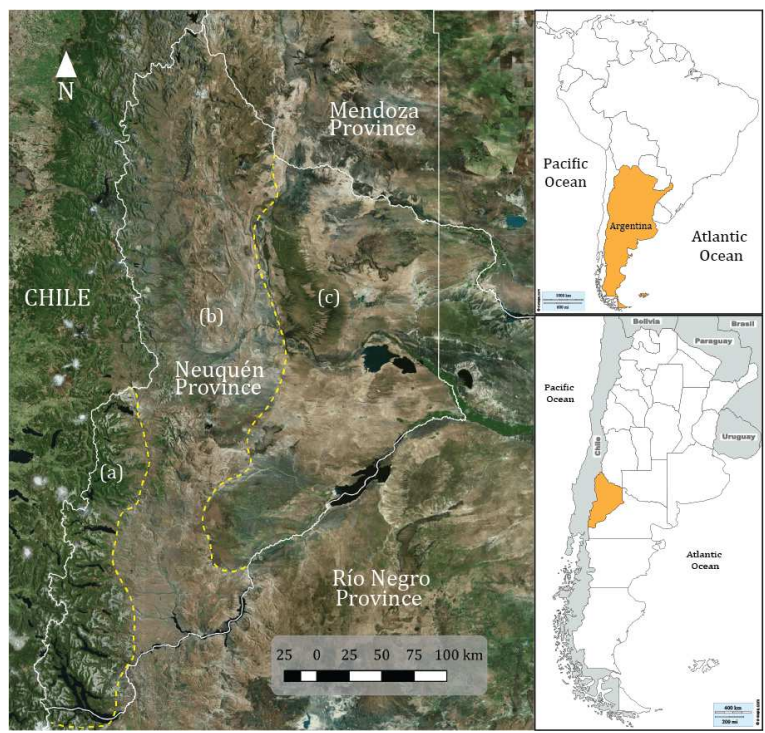}
\caption{The region of the present study, in the provinces of Neuqu\'en and R\'{\i}o Negro (Argentina). The three main environments are indicated: a) forest and lakes; b) ecotone and grass steppe; c) bush steppe.}
\label{patagonia}
\end{figure}

The present work is focused on the study and characterization of a set of rock art sites by means of network analysis. The region of study is located in the provinces of Neuqu\'{e}n and R\'{\i}o Negro (Argentina) \cite{vargas2020}  (Fig.~\ref{patagonia}). A previous version of this work was presented at the Archaeology Congress of Patagonia \cite{vargas2017} and is part of a larger project that aims to discuss the flow of information and mechanisms of cultural transmission in relation to archaeological hunter-gatherer models in northern Patagonia. We applied a novel analysis for rock art studies, based on communities detection, to solve some basic questions about the structure of the selected data and their spatial distribution. It is a first step towards the elucidation of the possible ways in which information and visual codes flowed in the past, and also to show how this gave place to the  specific distribution of the set over the landscape \cite{whallon2011}. A similar approach has been useful to model the geographical path of information, including but exceeding their topological dimension, which helped to reconstruct ancient hunter-gatherer paths, territories, action ranges, and other archaeological implications \cite{scheinsohn2011}. In turn, this information will serve as a comparative parameter to be contrasted with an archaeological model of land occupation and use of space in future research projects.

Our goal is to carry out an exploratory analysis at two different levels, drawn on shared features at each of the surveyed sites. Firstly, we evaluate the existence of correlations between the geographical distribution of the sites and the prevalence of different rock art motifs among them. Secondly, we perform a similar analysis but centered on the motifs themselves, in order to analyze the existence of well-defined ensembles of motifs. Based on the topological properties we intend to: a) detect the existence of highly relevant nodes, b) record how the nodes are organized, and c) check for the existence of communities by a modularity analysis. By means of such analysis we aim to unveil a meaningful aggregation of sites, linked by common cultural features that turns to be consistent with the possibility of cultural transmission or territorial behavior backed up by their geographical distribution.

At the same time that we presented the preliminary results of our analysis of community detection \cite{vargas2017}, a novel and significant article was published (Ref.~\cite{radivojevic18}, mentioned above), also applying this methodology but to a very different archaeological context. In their work, the authors evaluate the organization and interactions of copper using societies in the Balkans, from c.~6200 to c.~3200 BC, by means of a  community structure analysis of the  underlying  network  connecting several archaeological sites. The links among nodes were traced through the use of  chemical data of copper-based objects. The resulting division into communities revealed three dominant modular structures across the entire period, with  strong spatial and temporal significance. Later, during the review process of the present work, an additional paper appeared, applying the same methodology of community detection \cite{mazzucato2019}, also cited above. 

It is also worth mentioning that some of the sites studied here have been already analyzed in \cite{caridi16}, where the authors aim at  tracking cultural transmission paths by means of motif distribution in those sites. The analysis of the  constructed  mutual information network
between motifs allowed the authors  to propose a certain geographical organization with the presence of a hub region and satellite sites. The collection of sites presented here is larger, and the methodology is different as well.

\subsection {Archaeological background of northwestern Patagonia}

Rock art is one of the most abundant and relatively unknown archaeological evidence in our study region in northwestern Patagonia. These manifestations, either paintings or engravings on the rocks, constitute evidence of the communicative and symbolic dimension of the hunter-gatherer groups that inhabited this region. Rock art images occupied a central and active role in the transmission of information and knowledge for the immediate participants and also for their descendents.

In particular, the actual region of Neuqu\'{e}n and the northwestern R\'{\i}o Negro provinces, stands out for having a large number of sites with paintings and rock engravings, which are distributed in three different environments: the steppe, the ecotone and the forest (see Figure \ref{examples}). It has been established that they proliferated fundamentally during the last 3000 years, increasing both in number and in diversity of designs. This increase occurred in a context of demographic growth and possibly spatial  or territorial circumscription of the hunter-gatherer groups that inhabited these environments \cite{caridi16,crivelli2006,fiore2006}. In this sense, one of the most relevant characteristics of rock art in our study region is that, on a wide scale but also in short distances, there are different environments with an extensive set of rock art sites which have, in turn, differences in their visual and technical features. 

In the case of the steppe, more than 50 archaeological sites have been identified and studied with the presence of both paintings and engravings \cite{crivelli2006,boschin2009,vargas2019}. Historically these sites were studied by traditional methods of stylistic classification within a normative culture-history framework \cite{domingo2014}. In addition to this diffusionist explanation of social processes, these styles were used as chronological indicators to date other artwork. One of the most criticized aspects of these chronological styles was their lack of an explicit definition of the methodology used to formulate them, either qualitative or quantitative. Besides, there was also a lack of interest in studying the spatiality of rock art and its distribution over a large scale, as we are proposing here. In this way the styles were arbitrarily defined from the association, sometimes ambiguous and sometimes exclusive and rigid, of sets of motifs \cite{scheinsohn}. 

In synthetic terms, in our wide region of study three great rock styles were defined, the footprint style (\emph{estilo de pisadas})  \cite{menghin1957}, the fret style (\emph{estilo de grecas}) \cite{menghin1957}, also called Complex and Geometric-Abstract Trend (TAGC) \cite{gradin1999}, and the parallels style (\emph{estilo de paralelas}) \cite{menghin1957,schobinger1956,fernandez2000}. The temporal duration of these styles is not precise, because there is not direct dating of rock art. It has been possible, nevertheless, by dating archaeological layers covering some rock art motifs, to distinguish temporal blocks for the different repertoires of Patagonian rock art \cite{menghin1957,gradin1999,boschin}. In the case of northwestern Patagonia, the footprint style has been indirectly dated with an estimated antiquity of about 3000 years BP \cite{crivelli2006}, and was characterized by the representation of animals and humans through their footsteps or footprints, in conjunction with simple geometric designs such as circles, lines and points.  

In the case of the fret style or TAGC, it was defined as containing designs of considerable geometric complexity like labyrinths, staircase and crenellated motifs, a wide diversity of frets, crosses, squares and also framed and schematic human representations. At present, this style constitutes a challenge for archaeologists, since a great diversity of designs coexist within it, and there is no consensus about its extension in time and space. It is estimated, by indirect dating and the similarity of designs on artifacts found in archaeological stratigraphy, that the age of this style does not exceed 1000 years BP \cite{crivelli2006,albornoz2000,podesta2008}. 

In turn, the case of the parallels style has a different context, because in the archaeological literature it appears to be one of the unique and solitaire styles of the extreme north of the Neuqu\'{e}n province. Its age, also by no direct means, has been estimated in no more than 600 years BP \cite{fernandez2000}. 

In the ecotone and forest environments the rock art sites are located mainly near large bodies of water (Figure \ref{examples}). One of the peculiarities of these sites is that they contain entirely rock paintings, and their repertoire of designs differ markedly from the shapes of the neighboring steppe. In particular, there are designs that represent quadruped animals like huemuls (\emph{Hippocamelus bisulcus}) and guanacos (\emph{Lama guanicoe}), and also human figures that in many cases are represented with quadruped animals (interpreted as mounted horses). These characteristics have allowed archaeologists to postulate that this repertoire of motifs belongs to a more recent and different modality within the fret style, called the \emph{Modality of the Forest and Lake Area} or MABL \cite{albornoz2000}. Archaeologists argue that these paintings were made by a group substantially different from those of the steppe, that ethno-historic sources identified as an ethnic group called Puelches, which were fully adapted to this lacustrine-forested environment, since they 
dominated the use of boats and navigation \cite{hajduk2018}.

As can be seen, the presence of a large number and variety of rock art sites located in the different environments of Neuqu\'{e}n and the northwest of the R\'{\i}o Negro provinces, constitutes the evidence that diverse groups, during at least the last 3000 years, displayed substantially different visual, communicative and significant strategies. 


\section{Database and network construction}

\begin{figure}
\centering
\includegraphics[width=\columnwidth]{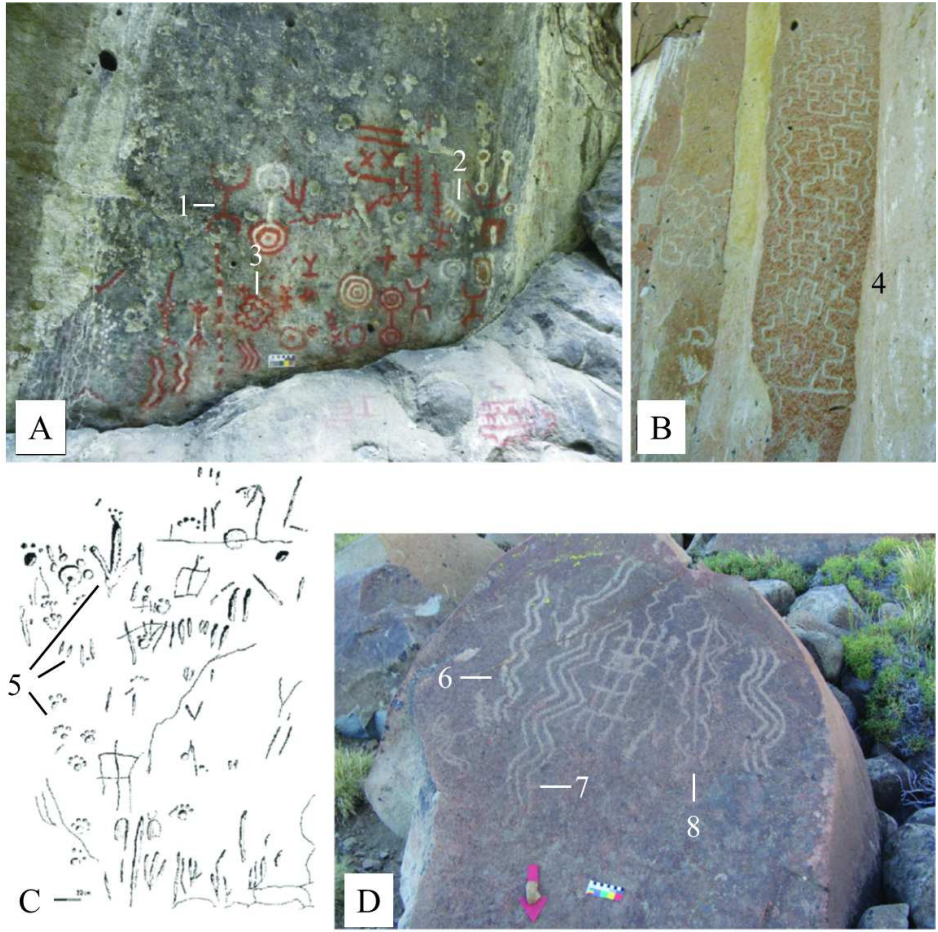}
\caption{Examples of typical motifs found in: A) Patagonian 
forest (photograph by F. E. Vargas, from \cite{vargasfoto}), B) and C) Patagonian steppe (photograph by M. T. Boschin, from \cite{boschin2009}), D) 
North of Neuqu\'{e}n province (photograph by A. Hajduk, from \cite{hajduk2009}). The motifs correspond to the following 
group classification referred to in the text: 1) anthropomorphic, 2) zoomorphic 
guanaco, 3) concentric fret, 4) fret (engraved), 5) footprints, 6) 
sinuous parallel lines, 7) zig-zag parallel lines, 8) axially symmetric figure.}
\label{examples}
\end{figure}

The database of prehistoric rock art consists of georeferenced sites, each of which contains paintings, engravings or both  \cite{vargas2017}. This 
relational database includes both published information about rock art sites from Patagonia as well as fieldwork data provided by our own team members 
and ARPat database \cite{vargas2017,fiore2006,fiore2018}. It contains information at the site scale and at the motif scale, which can be used 
independently or combined, according to digital queries defined by archaeological research questions. Out of this macro-regional database, we have 
selected for this paper 136 sites which regionally fall within Northwestern Patagonia. The quantity of motifs at each site ranges from 1 to 35 (with a mode of 7 motifs), totalizing 1095 motifs \cite{vargas2020}. Examples are shown in Fig.~\ref{examples}. Given the large variety of motifs designs, 
we classified them into groups, according to their geometric and morphology and also in terms of figurative an nonfigurative designs. This led to the 
definition of a total of 148 motif groups (see Appendix II). Examples of the groups are: concentric circles, parallel lines, feline tracks, frets, 
etc., some of which can be seen in Fig.~\ref{examples}.

The information about each site and the corresponding groups of motifs present 
in them was used to construct a bipartite network with sites and groups of motifs  on each side, 
as follows: a group of motifs is linked to a site if any of the motifs that belongs to the group is 
present at the site. From this bipartite network we built 
two different projections, with their nodes identified with only one of the two parts in each 
case. An example of the procedure is presented in Fig.~\ref{redes}. In this way 
we obtained the following two network types:

\begin{figure}
\centering
\includegraphics[width=0.8\columnwidth]{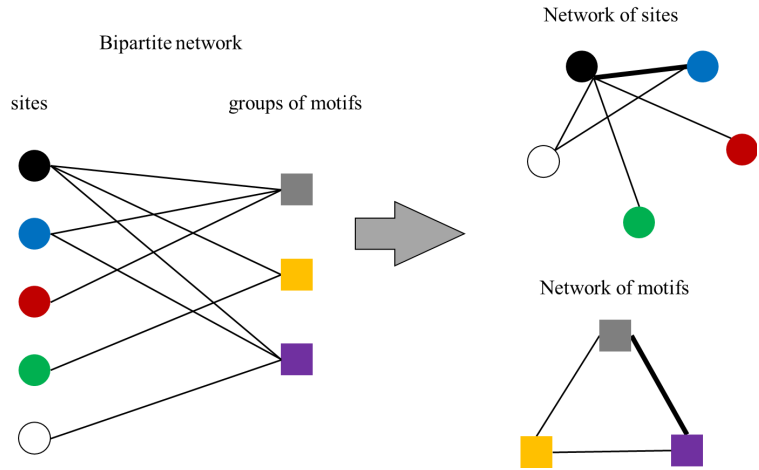}
\caption{Example of the construction of the networks. Thick lines correspond to stronger links, with the double of weight of the thin ones in the graph.}
\label{redes}
\end{figure}

\textbf{Network of sites.} In this case the nodes of the network are the archaeological sites. Two nodes 
are connected by a link if they share at least one group of motifs in common. If more than one 
group is shared, the link is weighted accordingly. It is important to note that the network 
thus constructed is an artificial mathematical object, which does not (necessarily) correspond to 
actual connections between the people that populated the sites and produced the art. 
This network is even blind with respect to the geographical location of the sites. Any 
topological structure present in the network will emerge from the complex sharing of groups of 
motifs between them.

\textbf{Network of motifs.} In this case the nodes are the categories to which each motif design 
corresponds, which as noted above, are called \emph{motif groups}. Two 
nodes of this network are linked if the corresponding motif groups appear at the same site. If a pair of groups appear 
together in more than one site, the link is weighted accordingly. 
This motifs network is complementary to the network of sites, and provides an insight 
on the possible relationships between the many motif categories of the set, thus shedding light on 
their stylistic classification.

These two types of networks were analyzed using standard tools of the theory of complex 
networks, in order to characterize their topological properties. Of particular interest in this 
study is the analysis of the partitions of the set of nodes, which define what are called 
\emph{communities} of the network. Social networks, built on contact between people, naturally 
break into communities that the mathematical analysis can detect. Bear in mind, however that our 
networks are not based on any known contact between people, but just by sharing common 
material culture elements instead. 

\section{Results}

\begin{table}
\caption{Summary of the topological properties of the networks. The columns show: number of 
vertices, density of links, mean path length, mean degree, global clustering coefficient, mean local 
clustering coefficient and mean betweenness centrality. The maximum betweenness centrality in each 
case corresponds to: site \#42 (Huechahue, $\beta=201.7$), group \#22 (anthropomorphic, 
$\beta=1161.7$).}
\label{tab:metrics}
\begin{tabular}{lcccccccc} 
Network  & $N$ &$\delta$ & $\langle L\rangle$ & $\langle k\rangle$ & $c$ & 
$\langle c\rangle$ & $\langle\beta\rangle$ \\
\hline
of sites & 136 & 0.51    & 1.75               & 68.2               & 0.72       
& 0.77               & 49.8  \\ 
of motifs& 148 & 0.13    & 2.33               & 19.6               & 0.37       
& 0.58               & 91.4  \\ 
\end{tabular}
\end{table}

Quantitative properties of the networks are summarized in 
Table~\ref{tab:metrics}. It is noticeable that the network of sites is rather dense, 
with a density of $0.51$ of all possible links. The average degree of sites is 
also large at $\langle k\rangle\approx 70$, from a distribution that, ranging 
from 1 to 119, is biased towards large values. We also 
see that the average path length of both networks is small, and that both are 
highly clustered. The higher clustering of the network of sites makes sense in the context of the 
spatial unity and cultural homogeneity (see Discussion).

The most interesting results are provided by the modularity analysis of both 
networks. This was done using two algorithms that support weighted graphs 
and that have proven to be adequate to this kind of social network \cite{yang2016}. The 
algorithm based on the leading  eigenvector of the 
community matrix~\cite{newman2006}, as well as the Louvain method 
\cite{blondel2008} were used. The resulting communities of the network of sites are shown 
in Fig.~\ref{comsites} (and listed in Appendix I), with the sites at their geographical location (which 
does not play any role in the calculation). The two algorithms give very similar 
results, with a small but relevant difference. The leading eigenvector method 
finds three communities of 
similar size (Fig.~\ref{comsites}, left). These will be referred  to according to their 
geographical range: North of Neuqu\'{e}n (black), Forest (green) and Steppe 
(red). It can be seen that, apart from a few sites between latitudes $-40$ and 
$-41$, the North of Neuqu\'{e}n community is the most geographically segregated 
of the three, occupying mainly the namesake region. The other two communities 
occupy mostly the whole range of the area under study, from the Andes 
mountain range on the west and over the Patagonian steppe at the center and 
east. The Steppe community, though, stays mainly east of the Andes. 

The Louvain algorithm finds four communities (Fig.~\ref{comsites}, right), with the overwhelming difference 
being the split of the Forest community found by leading eigenvector into two: a 
Nahuel Huapi community (green, concentrated around the southernmost tip of 
Neuqu\'{e}n, and corresponding to the Nahuel Huapi lake region), and a Valleys community 
(blue), extending along several river valleys in the transition region between 
forest and steppe. The size of the symbols in Fig.~\ref{comsites} corresponds to 
the betweenness centrality of the sites, $\beta$. We can see that the Valleys community 
is dominated by the sites with largest values of $\beta$ of the Forest community 
found by the leading eigenvector method. 

\begin{figure*}
\centering
\includegraphics[width=0.8\textwidth]{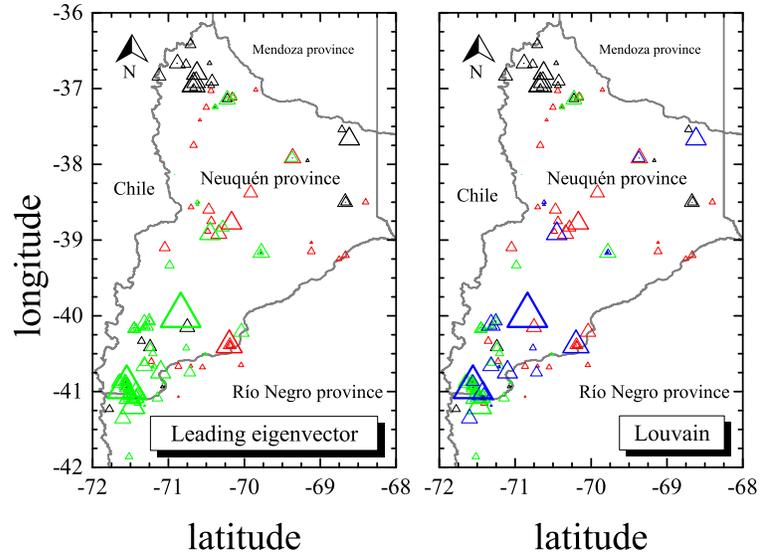}
\caption{Communities of the sites network, found by each algorithm (as indicated). Each color represents a community, 
and symbol size corresponds to the betweenness centrality of the site.}
\label{comsites}
\end{figure*}

Figure~\ref{comsites} is one of the main results of this paper. It is remarkable that these three 
(or four) homogeneous 
sets of sites are well segregated in geographical  space, even though the 
analysis, as mentioned, contains no information on the location of the sites. 
They show the potential existence of 
homogeneous blocks, with shared coherent information and visual codes, belonging to each one of this 
(mathematical) communities, which are based on a stylistic similarity given just by sharing 
groups of motifs. Indeed, the regions occupied by North of Neuqu\'{e}n province, Nahuel Huapi, Valleys and 
Steppe were inhabited by hunter-gatherer groups, which gradually occupied 
the environments in a discontinuous way during the last 4000 years. In this 
sense the network communities, despite not pondering the temporal 
variable of the data set, are coherent with some archaeological inferences about 
the occupation of the environments of northern Patagonia (see Discussion below). 

\begin{figure*}
\centering
\includegraphics[width=0.8\textwidth]{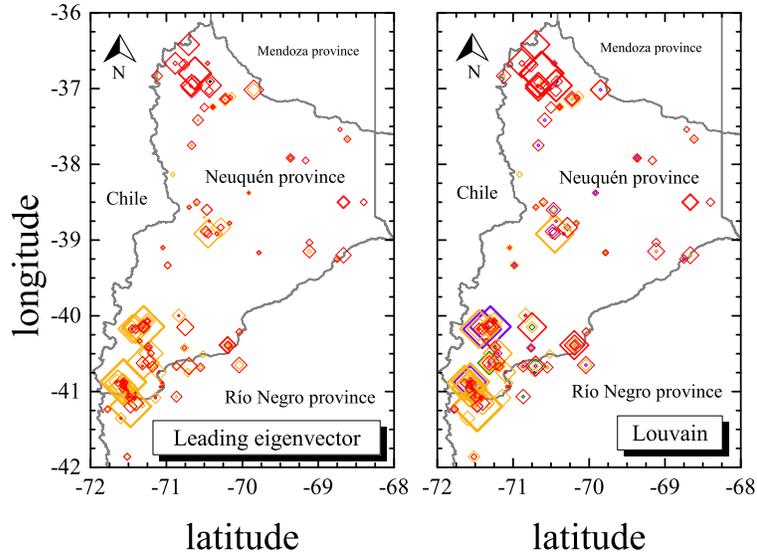}
\caption{Communities of the network of motifs, colored according to the module. Size 
corresponds to the quantity  of motifs found at each site, within each community. 
}
\label{comgroups}
\end{figure*}

We also analyzed the modularity structure of the network of groups of motifs. In this 
case, the communities indicate the existence of ensembles of motifs that tend to 
appear together, akin to stylistic modalities. In the following we will use the term ``module''
to refer to these communities of groups of motifs, especially to distinguish them 
from the ``communities'' of the network of sites. Again, bear in mind that the 
analysis is completely blind with respect to any other information about  the 
system. The modules found by the two algorithms are listed in corresponding 
tables in Appendix II. It is difficult to represent them on a map, because more 
than one thousand motifs are restricted to 136 sites. Figure~\ref{comgroups} is 
an attempt to do so. Each symbol is colored according to its 
module, and their size reflects the number of motifs of the corresponding group  present at 
the site. 

\begin{figure}
\centering
\includegraphics[width=0.6\columnwidth]{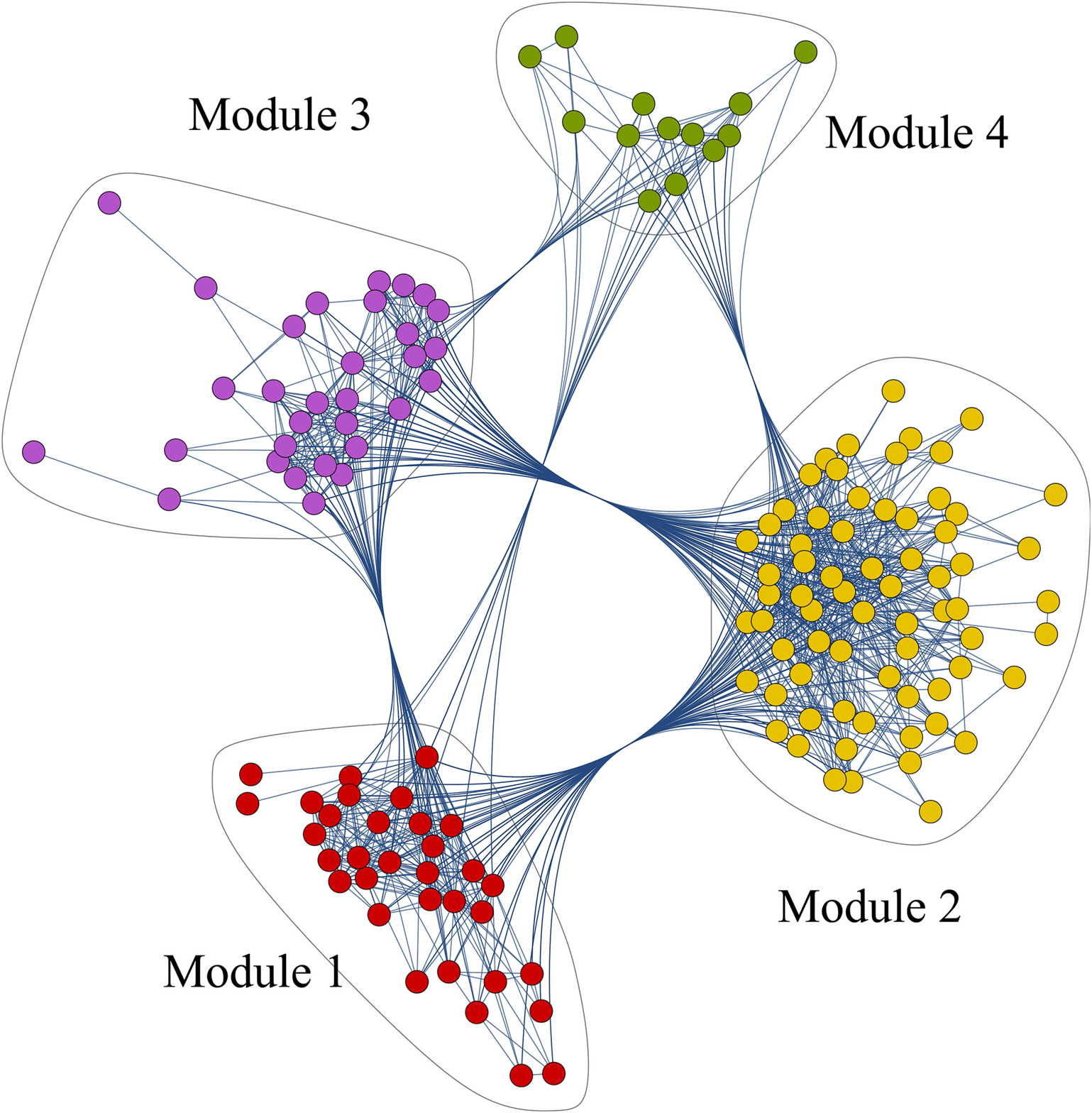}
\caption{Modules: communities of the motifs network (Louvain method).}
\label{comgroups2}
\end{figure}

In this case, the leading eigenvector algorithm  shows the existence of 2 modules 
(2 disconnected nodes, \emph{fig H} and \emph{fig rhombus}, were left out of the 
analysis). It is easy to see that these modules, just like the communities found 
in the network of sites, are also segregated in space. There is a Module 1 
(red), typical of the North of Neuqu\'{e}n, and a Module 2 (orange), typical of 
the South. Both of them mix, even at single sites, in the central valleys and 
steppe regions. The Louvain method again identifies subsets of the southern 
community. In this case the Module 2 of the leading eigenvector algorithm is 
split in three. The community structure is shown in Fig.~\ref{comgroups2}. 
Two of them (Modules 3 and 4, purple and green respectively) are smaller, and found mainly in the mixing regions of the steppe. 
We will discuss this structure and its relevance in the following section.

\section{Discussion}

Results are discussed here in the context of the actual system that the two identified networks represent. The modularity analysis carried out on the 
graphs has no information about the cultural, historical or geographical background of the sites and their motifs content. The resulting communities 
could have been just mathematical artifacts, devoid of anthropological meaning. However, we have found that both the sites network's communities and the 
motifs network's modules are full of relevant information. On the one hand, we see a clear distribution of the communities of sites over the landscape. On the other, the network of motifs allows us to start discussing not only their composition according to the classical styles classifications, but also a 
different configuration of those archaeological constructs.

Starting with the geographical distribution of the sites' communities, three (or four) distinct communities have been found: North of Neuqu\'{e}n, Forest 
(or Nahuel Huapi plus Valleys) and Steppe. These are easily recognizable as significant archaeological units. In the case of north of Neuqu\'{e}n, 
recently it has been shown by rock art and archaeological research ~\cite{vargas2019}, that the human groups that inhabited the North of Neuqu\'{e}n 
province participated actively in a regional dynamic of mobility and interaction together with the Maule region in Chile, at least during the last 
1000 years BP \cite{vargas2019,fernandez2000,hajduk2009,niemeyer1972}. Our modularity analysis was able to detect the context of production and use 
of rock art images that reinforces the particular identity of archaeological signals of the north of Neuqu\'{e}n apart from the rest of the province. 
Besides, the result of the motifs community analysis also shows an important presence of motifs that, under a classical framework, would be interpreted 
as belonging to the footsteps style (Apendix II, Louvain Module 1). As a consequence, the modularity analysis allows us to 
visualize a greater variability of motives than expected in the north of Neuqu\'{e}n, starting to reconfigure the status and distribution of the 
classical ``parallels style.''

As we briefly mentioned above, the human occupation of the ecotone and the steppe differs from that of the Andean forests, since 
it has more evidence of early human presence \cite{arias2012,crivelli1996,arias2010,arias2013,ceballos1982,crivelli1993,crivelli1996b,lanata1987} 
without the discontinuities observed in the forest environment and specifically in the Nahuel Huapi area \cite{hajduk2009}. The human groups that 
inhabited the ecotone and the steppe centered their economy and subsistence on guanaco (\emph{Lama guanicoe}) hunting, complemented sporadically with 
resources from the forest~\cite{lezcano2010}. It is also an extremely different habitat, desertic except in the rivers' valleys. In this case, the 
steppe environment, as well as the 
north of Neuqu\'{e}n, were mostly identified as the same communities by both algorithms (leading eigenvector and Louvain). Nevertheless, the results of the 
analysis show a clear distribution of one large community of sites, along the wide space of steppe, extending from north to south. The motifs of these sites are dominated by animal and human footprints with simple 
geometrical motifs, which is coherent with the archaeological background that postulates a minimum age of this rock art in 3000 years BP, produced with 
a particular engraving technique \cite{crivelli2006}. However, although it contains many geometrical patterns (such as frets), they are very different 
from the similar geometrical patterns found in forests. Again, our analysis was able to identify the steppe as one of the unique communities with a 
great diversity of motifs, mixing the classical footstep style with the motif commonly adscribed to the fret style. 

The case of the Forest community (leading eigenvector) deserves special attention, since it was split by Louvain into two communities (Nahuel 
Huapi and Valleys): this has archaeological implications. From a biogeographic perspective, the southern half of Neuqu\'{e}n has a phytogeographic 
transition: the ecotone between the Patagonian forest and the steppe. It was revealed by the Louvain method that it has differences in the communities of 
sites and motifs, as if it were a transitional region. As mentioned, previous archaeological approaches to rock art have proposed that the forest and 
lake environments of southern Neuqu\'{e}n and the west of the provinces of R\'{\i}o Negro (especially from the Manso river to the Nahuel Huapi and Lacar lakes) 
possess only one and unique style: the style of frets, called \emph{Modality of the Forest and Lake Area} 
\cite{albornoz2000,hajduk2018}. This style was 
related to hunter-gatherer groups with aquatic mobility, well adapted to lacustrian environments, and it has been used as an archaeological indicator 
to mark the differences with hunter-gatherer groups with only pedestrian mobility that inhabited the ecotone and the neighboring steppe 
\cite{hajduk2018}.

Even if the fret style is ubiquitous in both environments, there are several differences in terms of the morphology, techniques and colors of their 
composing motifs. In the former case, the basin of Manso river, Nahuel Huapi and Lacar lakes, fret style motifs are usually monochromatic red and 
mostly made with painting techniques \cite{albornoz2000,podesta2008,hajduk2018,albornoz1996}. There is a transition to polychromy in the valley of the 
Traful river \cite{silveira1996,silveira2014} and also the Meliquina river 
and lake, near the ecotone environment, and they are definite polychromous in the steppe \cite{vazquez2010}. This difference could be explained not only 
from a point of view of the availability of raw materials, because the Nahuel Huapi sites are relatively close to the steppe where raw materials to 
make colors are available, but also as an intentional cultural choice with the purpose of marking the landscape differently. It is remarkable that, 
even though all this cultural and rock art information was not loaded into the database used to build the networks, the modularity analysis 
was able to detect these subtle differences and find these two (sub-)communities.

Another interesting result of the split into two communities of the southern half of Neuqu\'{e}n is the presence of nodes (sites) with large betweenness 
centrality (the size of the symbols in Fig.~\ref{comsites}), which also has possible archaeological implications. In particular, the Huechahue site has 
the highest centrality and it is located precisely at the ecotone region in one of the valleys that connects the Andes with the steppe. As we saw, 
this is both a biogeographic transition and a boundary between site communities. This high centrality measure of the Huechahue site could be a 
consequence not only of its geographically strategical position but also of its probable role as an aggregation site, or a place where hunter-gatherer 
sub-groups met at specific times of the year (e.g. seasonally, annually, etc.).

Regarding the network of motifs, the modularity analysis also produced relevant communities that can be identified as a set of coherent information on 
spaces where the information circulated in a relatively standardized way. This automatic procedure gives independent support to analyze modalities, and 
proposes the possible existence of different ones within the fret style, about which there is an ongoing controversy 
~\cite{scheinsohn,boschin}. In this case, we were able to distinguish qualitatively and quantitatively the differences expressed between the motifs 
found in the forest and those belonging to the ecotone and steppe environment. In the case of the first (see Appendix II), we verified that the 
presence of anthropomorphs and zoomorphs is a clear signal that characterizes the forest's rock art. In addition, Module 2 presented more variety of 
geometrical motifs than expected for its environment, because the stylistic modality of the forest was defined as consisting predominantly of 
figurative motifs \cite{hajduk2009}. This allows us to attempt a redefinition of the repertoire of motifs belonging to the traditional Modality of the 
Forest and Lake Area. However, it should be noted that this difference may also be due to a sample bias, not only because we focused on the western side 
of Neuqu\'{e}n province, but also because the forest environment has better records and more publications of archaeological sites. On the latter (see 
Appendix II, Modules 3 and 4), we identified and verified that the fret motifs are more varied and complex, as well as accompanied by simple geometric and 
figurative motifs. This implies a greater diversity, possibly due to the differences observed in terms of the history of human occupation of the 
steppe and the ecotone during the last 3000 years BP.

In summary, our community detection analysis shows a clear differentiation of networks of sites with no overlapping between them and distributed over 
three different environments. This scenario can be interpreted as a possible stage during the late Holocene where different groups were marking and 
using each landscape in a different way. However, taking into account that in the database we have integrated and analyzed sites covering a temporal range of 
about 3000 years, these networks that we observe distributed in space could be the product of different groups marking the landscape in 
different periods of time. So, there is a possibility that we were seeing transient networks. Nevertheless, we think that the analysis is robust because, 
despite of the temporal problem, we were able to detect just three or four networks, either by the Louvain or leading eigenvector algorithms. In a synthetic 
way, it is clear from independent evidence that the steppe environment, including the north of Neuqu\'{e}n, are more connected to each other than between them and the sites of the 
ecotone and forest environments, which in turn split in at least two more communities. All these networks are compatible with the archaeological 
background in terms of the human groups that inhabited these diverse environment, and also with the expectation from the possible demographic density in a human occupation scenario of these spaces. 

In order to test the significance of our findings regarding the community structure of the networks, we have randomized the original networks according to the following procedure. We chose two pairs of connected nodes and interchange their links, avoiding double links and loops. The repeated application of this procedure randomizes the network preserving the degree of each node. By controlling the number of such changes, we can control the randomization introduced in the original network. We have observed that the modularity of these networks decreases with the randomization and, correspondingly, the number of communities increases. Also that their community structure lacks any relevant information content from an anthropological point of view, since the links are progressively randomized. The results of our analysis, instead, can be framed in the existing body of archaeological knowledge, as discussed.

\section{Conclusion}

The application of network analysis to the rock art database of northern 
Patagonia brings new insight in two important ways. Firstly, we want to 
highlight the methodological aspect of community detection as a formal method to 
identify possible anthropological and archaeological processes at different 
spatial scales \cite{radivojevic18,mazzucato2019}. This methodological innovation allows us to visualize and 
discriminate clusters of sites and motifs in a suprarregional scale, which is not 
a common approach in rock art studies of Patagonia. 

The detection of communities in rock art studies could serve as a first 
step to discover in which manner the repertoire of sites and motifs are 
distributed and linked through the landscape, and which sites could have particular 
properties or roles within the network. In our case, the 
communities were coherent with the archaeological background of hunter-gatherers of 
northern  Patagonia. This type of analysis will allow us to generate 
hypotheses about the relevance of certain sites or sets of sites within the 
network, to be contrasted with other lines of evidence and archaeological 
models of land use, mobility and subsistence.  

Finally, the most challenging and promising issue that network analysis of rock art can provide is the detection of the variability of the rock art 
record, which cannot be visualized or identified with traditional archaeological methods. Such is the case of those sites or group of sites which do 
not fit into the regional styles or repertoires previously established and with traditional consensus in the academic arena. As was mentioned by 
Ta\c{c}on~\cite{tacon2013}, we need to pay attention to those sites that are in between two or more well-known regional styles, in order to achieve a 
more accurate interpretation on their role in past hunter-gatherer systems. The relational approach of network analysis and specifically of community 
detection is a fruitful path to explore, as well as their implication into regional archaeological debates.

\section*{Funding}
This work was supported by Agencia Nacional de Promoci\'on Cient\'{\i}fica y Tecnol\'ogica
[PICT-2014-1558]; Universidad Nacional de Cuyo [06/506]. 

\section*{Acknowledgements}
F.E.V. acknowledges CONICET for the partial support of this research through a doctoral fellowship.

\section*{Appendix I}

Communities of the network of sites, according to the leading eigenvector and 
the Louvain methods. The sites are derived from a compilation of bibliography 
and new surveys performed by one of the authors (F. E. Vargas), and are 
described in full in~\cite{vargas2020}.

\medskip
\hrule
\medskip
\noindent Leading eigenvector method.
\begin{description}
\item[\textbf{North of Neuqu\'{e}n:}] Abrigo de las Mosquetas, 
Abrigo de las Torcazas, 
Agua escondida,
Alero Coliguay,
Butal\'{o}n Norte,
Caj\'{o}n de Flores,
Ca\~{n}ada de las Minas,
Casa de Piedra,
Caverna de los Gatos,
Cerro de las Brujas,
Colomichico (E y G),
Corral 1,
El Chacay,
Estancia Jones,
Las Chaquiras,
Los Barriales 1,
Los Barriales 2,
Los Radales,
Mata Molle,
Mogotillos Arriba,
Molulco-Mogotillos,
Pampa Linda,
Parque Diana,
Paso Valdez,
Piedra Bonita,
Pozo del Loro,
Puesto Marchan,
Rinc\'{o}n de Las Papas.
\item[\textbf{Steppe:}] 
Arroyo Llano Blanco,
Aguada del Carrizal,
Alero de la Vizcacha,
Alero La Marcelina,
Alero La Pava,
Alero Larriviere,
Ca\~{n}ad\'{o}n de Santo Domingo,
Cancha Huinganco,
Casa de Piedra Ortega,
Cerro Nonial,
CH1-CNG1-CNG3-CY-PC5-PC6-PdT2-PdT3-PdT4,
Choc\'{o}n Chico,
Cueva del Ca\~{n}ad\'{o}n de la Piedra Pintada,
Cueva Epull\'{a}n Chica,
Cueva Epull\'{a}n Grande,
Cueva Visconti,
El Manantial,
La Medialuna,
La Oquedad,
Los Chenques II,
Los Grabados,
Los Rastros,
Malal Huaca o Arroyo Mala Vaca,
Pared\'{o}n Las Lajitas,
Pe\~{n}a Haichol,
Piedra Pintada,
Piedra Pintada del Manzanito,
Planicie Banderita,
Portada Covunco,
Puerto Tranquilo IX,
Puesto Bustingorria,
Quili Malal.
\item[\textbf{Forest:}]
Abra Ancha,
Abra Grande,
Abrigo de las Cruces,
Abrigo de Media Falda,
Abrigo del Ciervo,
Abrigo del Risco,
Abrigo Loma Alegre,
Agua del Pino,
Alero las Mellizas,
Alero Los Cipreses,
Alero Maqui,
Aleta (Puesto Mu\~{n}oz),
Antepuerto (iv 8),
Arroyo Seco,
Bah\'{\i}a L\'{o}pez,
Bajada de la Arena,
Campanario 2,
Casa de Piedra de Curapil,
Catritre 1,
Cementerio R\'{\i}o Limay,
Cerro Abanico,
Cerro Carb\'{o}n,
Cerro El Huecu,
Cerro Leones,
Chacay Melehue I,
Chacay Melehue II,
Chacay Melehue IV,
Chape,
Chenque Pehu\'{e}n,
Cueva Alihu\'{e}n,
Cueva Picaflor,
Cuevas Gaudianas,
Curruhuinca 1,
El Chenque 1,
El Chenque 2,
El Chenque M de la Barda Negra,
El Crucero,
El Monito,
El Tr\'ebol,
Estancia Chacabuco,
Extremo SO (iv 10),
Gingins,
Huechahue,
La Ramadilla,
Lago Guillelmo,
Lago Moreno,
Laguna del Pescado (iv 7),
Los R\'{a}pidos
Nariz del Diablo 1,
Nichos 1 y 2,
Norte de Puerto Pampa (iv 5),
Norte de Puerto Pampa (iv 6),
Norte de Puerto Vargas (iv 3),
Pared\'{o}n Bello,
Pared\'{o}n con arte (PTA 4),
Piedra del Maqui,
Piedra Trompul,
Potrero de la Bah\'{\i}a,
Puente de Tierra (iv 9),
Puerto Anchorena (iv 8a),
Puerto Chavol 1,
Puerto Chavol 2 (Nariz del Diablo II sensu Pedersen),
Puerto Tigre,
Puerto Tranquilo I (iv 2a),
Puerto Tranquilo III (iv 2),
Puerto Tranquilo VI,
Puerto Tranquilo VII,
Puerto Vargas (iv 4),
Punta Verde (iv 1),
Quebrada de la Piedra Pintada,
Queutre-Inalef,
Quila Quina 1,
Rinc\'{o}n Chico,
R\'{\i}o Minero I,
R\'{\i}o Minero II,
Villa Coihues.
\end{description}

\medskip
\hrule
\medskip
\noindent Louvain method.

\begin{description}
\item[\textbf{North of Neuqu\'{e}n:}]
Abrigo de las Mosquetas,
Abrigo de las Torcazas,
Abrigo del Risco,
Alero Coliguay,
Butal\'{o}n Norte,
Caj\'{o}n de Flores,
Ca\~{n}ada de las Minas,
Casa de Piedra,
Caverna de los Gatos,
Cerro de las Brujas,
Colomichico (E y G),
Corral 1,
El Chacay,
Estancia Jones,
Laguna del Pescado (iv 7),
Las Chaquiras,
Los Barriales 1,
Los Barriales 2,
Los Radales,
Mogotillos Arriba,
Molulco-Mogotillos,
Pampa Linda,
Paso Valdez,
Piedra Bonita,
Pozo del Loro,
Puesto Marchan,
Rinc\'{o}n de Las Papas.
\item[\textbf{Steppe:}]
Arroyo Llano Blanco,
Aguada del Carrizal,
Alero de la Vizcacha,
Alero La Marcelina,
Alero La Pava,
Alero Larriviere,
Ca\~{n}adon de Santo Domingo,
Cancha Huinganco,
Casa de Piedra Ortega,
Cerro Nonial,
CH1-CNG1-CNG3-CY-PC5-PC6-PdT2-PdT3-PdT4,
Chacay Melehue I,
Choc\'{o}n Chico,
Cueva del Ca\~{n}ad\'{o}n de la Piedra Pintada,
Cueva Epull\'{a}n Chica,
Cueva Epull\'{a}n Grande,
Cueva Visconti,
El Manantial,
La Medialuna,
La Ramadilla,
Los Chenques II,
Los Grabados,
Los Rastros,
Malal Huaca o Arroyo Mala Vaca,
Mata Molle,
Pared\'{o}n Las Lajitas,
Parque Diana,
Pe\~{n}a Haichol,
Piedra Pintada,
Piedra Pintada del Manzanito,
Planicie Banderita,
Portada Covunco,
Puerto Tranquilo IX,
Puesto Bustingorria,
Quili Malal,
Rinc\'{o}n Chico,
Villa Coihues.
\item[\textbf{Nahuel Huapi:}]
Abra Ancha,
Abrigo de las Cruces,
Abrigo de Media Falda,
Abrigo Loma Alegre,
Alero las Mellizas,
Aleta (Puesto Mu\~{n}oz),
Antepuerto (iv 8),
Bah\'{\i}a L\'{o}pez,
Campanario 2,
Catritre 1,
Cementerio R\'{\i}o Limay,
Cerro Abanico,
Cerro Leones,
Chacay Melehue II,
Chacay Melehue IV,
Chape,
Chenque Pehu\'{e}n,
Cueva Picaflor,
Cuevas Gaudianas,
Curruhuinca 1,
El Chenque M de la Barda Negra,
El Crucero,
El Monito,
El Tr\'{e}bol,
Extremo SO (iv 10),
Gingins,
Lago Guillelmo,
Lago Moreno,
Nariz del Diablo 1,
Nichos 1 y 2,
Norte de Puerto Pampa (iv 6),
Norte de Puerto Vargas (iv 3),
Piedra del Maqui,
Piedra Trompul,
Puente de Tierra (iv 9),
Puerto Anchorena (iv 8a),
Puerto Chavol 1,
Puerto Chavol 2 (Nariz del Diablo II sensu Pedersen),
Puerto Tigre,
Puerto Tranquilo I (iv 2a),
Puerto Tranquilo VI,
Puerto Tranquilo VII,
Puerto Vargas (iv 4),
Punta Verde (iv 1),
Queutre-Inalef,
Quila Quina 1,
R\'{\i}o Minero I.
\item[\textbf{Valleys:}]
Abra Grande,
Abrigo del Ciervo,
Agua del Pino,
Agua Escondida,
Alero Los Cipreses,
Alero Maqui,
Arroyo Seco,
Bajada de la Arena,
Casa de Piedra de Curapil,
Cerro Carb\'{o}n,
Cerro El Huecu,
Cueva Alihu\'{e}n,
El Chenque 1,
El Chenque 2,
Estancia Chacabuco,
Huechahue,
La Oquedad,
Los R\'{a}pidos,
Norte de Puerto Pampa (iv 5),
Pared\'{o}n Bello,
Pared\'{o}n con arte (PTA 4),
Potrero de la Bah\'{\i}a,
Puerto Tranquilo III (iv 2),
Quebrada de la Piedra Pintada,
R\'{\i}o Minero II.
\end{description}

\section*{Appendix II}

Communities of the network of motifs, called ``modules'' in this work, according to the leading eigenvector and 
the Louvain methods. The sites are derived from a compilation of bibliography 
and new surveys performed by one of the authors (F. E. Vargas), and are 
described in full in~\cite{vargas2020}.

\medskip
\hrule
\medskip
\noindent Leading eigenvector method (fig rhombus and fig H disconnected and unclassified).

\begin{description}
\item[\textbf{Module 1:} Simple geometric, zoomorph, hand- and footprints.]
circle,
concentric circles,
associated circles,
complex triangle+circles+rhombi assoc,
stairlike,
indet combined strokes,
fig L,
axial sym,
phytomorphic,
straigh line+short trans append,
parallel lines,
winding parallel lines,
zig-zag parallel lines,
straight lines,
segmented lines,
sinuous lines,
zig-zag,
hands,
irreg oval,
feline footprint,
guanaco footprint,
human footprint,
points,
grouped points,
aligned points,
reticulate,
rhombi,
triangles,
aligned triangles,
tridigit,
zoomorphic bird,
zoomorphic quadruped,
indet zoomorphic.
\item[\textbf{Module 2:} Antropomorph, complex geometric.]
anthropomorphic,
anthropo-zoomorphic,
finger slide,
finger+palm slide,
bidigit,
circle+lines,
circle+appendix,
side-by-side circles,
side-by-side circles+appendix,
circles+radial appendices,
side-by-side conc circles,
conc circles+appendix,
conc circles+inside element,
joined circles,
circle+inside elment,
clepsydra,
clepsydra+inside element,
cruciform,
cruciform+appendices,
cruciform+inside\&outside elements,
grouped cruciform,
conc cruciform,
double cruciform,
stairlike cruciform,
irreg cruciform,
triple cruciform,
joined cruciform,
squares,
side-by-side squares,
squares+inside element,
segmented squares,
framed,
irreg framed,
zig-zag framed,
spiral+geometric,
star,
arc,
arc+inside element,
square+inside element,
concentric squares,
fig stairlike,
geometric indet,
fig I,
fig curved lines,
fig horiz+vert lines,
fig broken lines,
fig 8,
fig 8+inside element,
fig orthog stairlike,
pentagonal,
straight line+appendices,
straight lines indet,
sinuous fig+inside elements,
fig T,
fig trapezoidal,
fig triangular,
triang rhomboidal fig,
fig U,
fig V,
fig invert Y,
linear crenelated,
linear X-like,
linear Y-like,
fret,
double fret,
stairlike fret,
irreg fret,
irreg fret+inside element,
dashes,
hole,
hole+lines,
rider,
labyrinth,
stairlike labyrinth,
irreg labyrinth,
square labyrinth,
angled lines,
curved lines,
curved lines+radial appendices,
stairlike lines,
orthogonal lines,
orthogonal+curved lines,
broken lines,
straight lines+dashes,
sinuous lines w appendix,
spots,
eye-like
oval,
oval+appendices,
oval+inside element,
oval+lines,
pointlike circular,
dots,
rectangles,
rectangles w appendices,
rectangles w inside elements,
conc rectangles,
filled rectangles,
segmented rectangles,
rhombi w appendices,
stairlike rhombi w inside elements,
segmented rhombi,
semicircle,
triangles+circle,
side-by-side triangles,
stairlike triangles,
cruciform stairlike triangles,
opposite triangles,
tridigit+circles+appendices,
zoomorphic+stairlike,
zoomorphic frog,
zoomorphic guanaco,
zoomorphic snake.
\end{description}

\medskip 
\hrule
\medskip
\noindent Louvain method (fig rhombus and fig H disconnected and unclassified).
\begin{description}
\item[\textbf{Module 1:} Simple geometric, zoomorph, hand- and footprints.]
circle,
stairlike,
indet combined strokes,
fig L,
axial sym,
phytomorphic,
square labyrinth,
straigh line+short trans append,
orthogonal+curved lines,
parallel lines,
winding parallel lines,
zig-zag parallel lines,
straight lines,
segmented lines,
sinuous lines,
zig-zag,
hands,
oval+appendices,
oval+inside element,
oval+lines,
feline footprint,
guanaco footprint,
human footprint,
points,
grouped points,
aligned points,
rectangles w appendices,
reticulate,
rhombi,
triangles,
aligned triangles,
tridigit,
indet zoomorphic.
\item[\textbf{Module 2:} Antropomorph and complex geometrical.]
anthropomorphic,
anthropo-zoomorphic,
finger slide,
finger+palm slide,
side-by-side circles+appendix,
concentric circles,
circles+radial appendices,
clepsydra,
clepsydra+inside element,
complex triangle+circles+rhombi assoc,
cruciform,
cruciform+appendices,
double cruciform,
stairlike cruciform,
irreg cruciform,
triple cruciform,
squares,
side-by-side squares,
squares+inside element,
segmented squares,
framed,
irreg framed,
spiral+geometric,
star,
arc,
arc+inside element,
square+inside element,
geometric indet,
fig I,
fig 8,
fig orthog stairlike,
sinuous fig+inside elements,
fig trapezoidal,
fig triangular,
triang rhomboidal fig,
fig V,
fig invert Y,
linear crenelated,
linear X-like,
frte key,
double fret,
irreg fret,
dashes,
rider,
labyrinth,
stairlike labyrinth,
curved lines,
curved lines+radial appendices,
stairlike lines,
orthogonal lines,
broken lines,
straight lines+dashes,
spots,
oval,
pointlike circular,
dots,
rectangles,
rectangles w inside elements,
filled rectangles,
segmented rectangles,
stairlike rhombi w inside elements,
triangles+circle,
side-by-side triangles,
stairlike triangles,
cruciform stairlike triangles,
opposite triangles,
zoomorphic+stairlike,
zoomorphic bird,
zoomorphic quadruped.
\item[\textbf{Module 3:} Complex circles and undulated.]
bidigit,
circle+appendix,
side-by-side circles,
circles+radial appendices,
conc circles+appendix,
conc circles+inside element,
cruciform+inside\&outside elements,
joined cruciform,
concentric squares,
fig stairlike,
fig horiz+vert lines,
fig broken lines,
fig 8+inside element,
pentagonal,
straight line+appendices,
straight lines indet,
fig T,
fig U,
stairlike fret,
irreg fret+inside element,
irreg labyrinth,
angled lines,
sinuous lines w appendix,
eye-like
irreg oval,
rhombi w appendices,
segmented rhombi,
semicircle,
tridigit+circles+appendices,
zoomorphic frog,
zoomorphic snake.
\item[\textbf{Module 4.}]
Circle+lines,
side-by-side conc circles,
joined circles,
associated circles,
grouped cruciform,
conc cruciform,
zig-zag framed,
fig curved lines,
linear Y-like,
hole,
hole+lines,
conc rectangles,
zoomorphic guanaco.
\end{description}

\end{document}